# On the lipid-bacterial protein interaction studied by quartz crystal microbalance with dissipation, transmission electron microscopy and atomic force microscopy

**Mihaela Delcea,[1] Susana Moreno-Flores,[1] Dietmar Pum,[2] Uwe Bernd Sleytr,[2] and José Luis Toca-Herrera[1,*]**

[1]Biosurfaces Unit, CIC BiomaGUNE, Paseo Miramón 182, 20009 San Sebastián, Spain

[2]Center for NanoBiotechnology, BOKU - University of Natural Resources and Applied Life Sciences Vienna, Gregor-Mendel-Strasse 33, 1180 Vienna, Austria, Europe

*Corresponding author:

José L. Toca-Herrera: jltocaherrera@cicbiomagune.es

**Abstract**

The interaction between the bacterial S-protein SbpA on different types of lipid membranes has been studied using atomic force microscopy, transmission electron microscopy, and quartz crystal microbalance with dissipation.

On one hand, It has been found that the bacterial forms two dimensional nanocrystals on zwitterionic DOPC bilayers and negatively charged DMPG vesicles adsorbed on mica, on zwitterionic DPPC and charged DPPC/DMPG (1:1) monolayers adsorbed on carbon grids.

On the other hand, SbpA protein adsorption took place on zwitterionic DOPC bilayers and DOPC/DOPS (4:1) bilayers, previously adsorbed on silicon supports. SbpA adsorption also took place on DPPC/DOPS (1:1) monolayers adsorbed on carbon grids.

Finally, neither SbpA adsorption, nor recrystallization was observed on zwitterionic DMPC vesicles (previously adsorbed on polyelectrolyte multilayers), and on DPPC vesicles supported on silicon.

**Introduction**

Phospholipids represent the major component of biological membranes. Therefore, it is not strange that in the last decades phospholipids has been a subject of great interest in biochemistry, chemistry and polymer science.

Due to their amphiphilic nature, when a suspension of phospholipids is mechanically dispersed in aqueous solution several types of structures can be formed: i) phospholipid vesicles [1, 2]; ii) phospholipid monolayers [3-6]; iii) free standing foam films, formed by two monolayers of surfactant molecules [7,8] and, iv) supported phospholipid bilayers [9, 10].

These model systems are particularly interesting when they are associated with proteins [11, 12]. Thus, lipid bilayers have been extensively used as matrix for protein adsorption or insertion, some examples are annexin A5 [13], streptavidin [14], histidine-tagged (His-tagged) membrane proteins [15]. So, different approaches and molecules have been used to functionalize a lipid monolayer; for example: biotinylated amphiphile-streptavidin system [16], protein A [17], and lipid linkers [18].

Bacterial surface layer proteins (S-layers) are two-dimensional protein lattices forming the outermost cell envelope component in a broad spectrum of bacteria and archaea [19-20]. They are composed of a single protein or glycoprotein species and exhibit either oblique, square or hexagonal lattice symmetry with center-to-center units in the range 3-30 nm. Their thickness oscillates between 5 and 10 nm, presenting pores of uniform morphology and size in the range 2-8 nm [21]. Especially interesting for bionanotechnological applications is the ability of S-layer proteins to recrystallize onto lipid monolayers [22-24], solid-supported membranes [25, 26] or liposomes [27, 28]. We should not forget that S-layer supported lipid membranes have been optimized for billions of years of evolution in most extreme habitats. Thus, they are biomimetic structures mimicking the supramolecular building principles of archaeal cell envelope, being promising candidates for structure-function studies on molecular nanotechnology.

In this work, saturated and unsaturated lipids with different head groups and charge have been used to study the interaction of the SbpA protein from *Lysinibacillus sphaericus* CCM 2177 (former *Bacillus sphaericus*) with liposomes, lipid bilayers and monolayers. We have also studied the influence of lipid mixtures on SbpA recrystallization. In particular, the lipids used were: phosphatidylcholine (PC), phosphatidylserine (PS) and phosphatidylglycerol (PG). It is known that phosphatidylcholine (PC) and phosphatidylserine (PS) phospholipids are required for normal cellular structure and function. On one hand, saturated PC (phosphatidylcholine) phospholipids have gained the interest in the pharmaceutical branch because they promote metabolism through the cell membrane. On the other hand, unsaturated PC lipids might be important in the formation of a lipid reservoir, i.e. in the initial adsorption of lipids to the interface or in the regulation of surface tension during the respiratory cycle. Since PC lipids are uncharged, we have also used an anionic phospholipid, phosphatidylglycerol (PG), which is one of the major membrane phospholipids and assist in translocation of proteins across membranes and initiation of DNA replication [29-31].

In addition, a more complex biomimetic supramolecular structure consisting of the following steps: i) polyelectrolyte multilayers (PEM) on silicon, ii) 1,2-dimyristoyl-sn-glycero-3-phosphocholine (DMPC) on PEM and iii) SbpA protein adsorption on DMPC, have been also investigated.

**Materials and Methods**

*Lipids:* All lipids used in this work were purchased from Avanti Lipids. The corresponding structures and transition temperatures (Tm) are shown in Figure 1.

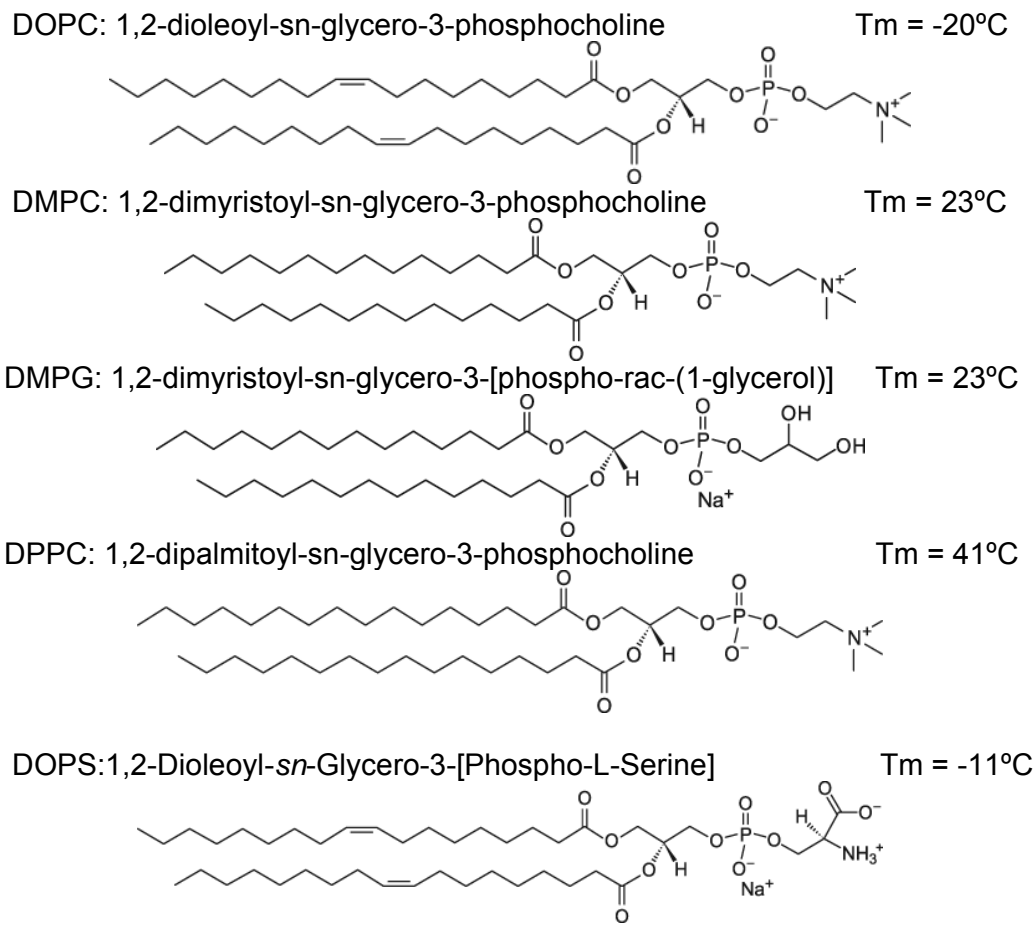

DOPC: 1,2-dioleoyl-sn-glycero-3-phosphocholine    Tm = -20°C

DMPC: 1,2-dimyristoyl-sn-glycero-3-phosphocholine    Tm = 23°C

DMPG: 1,2-dimyristoyl-sn-glycero-3-[phospho-rac-(1-glycerol)]    Tm = 23°C

DPPC: 1,2-dipalmitoyl-sn-glycero-3-phosphocholine    Tm = 41°C

DOPS: 1,2-Dioleoyl-*sn*-Glycero-3-[Phospho-L-Serine]    Tm = -11°C

Figure 1. Chemical structure and transition melting temperature (Tm) of the lipids used in this work.

*Polyelectrolytes:* Poly(ethylenimine) (PEI, Mw = 750 kDa), Poly(sodium 4-styrenesulfonate) (PSS, Mw = 70 kDa), Poly(allylamine hydrochloride) (PAH, Mw = 70 kDa) were obtained from Sigma-Aldrich (Munich, Germany) and were used as received. PEI and PAH are polycationic and PSS is polyanionic.

*Buffer solutions:* Two buffer solutions were used in this work: Tris buffer (0.5 mM Tris-HCl, 10 mM $CaCl_2$, pH 9), and HEPES buffer (10 mM HEPES, 2 mM $CaCl_2$, 150 mM NaCl, pH 7.4).

*Substrates:* 5 MHz QCX303 silicon dioxide coated quartz crystals (Q-Sense AB, Gothenburg, Sweden) were cleaned before surface preparation by immersion in a 6:1:1 (vol/vol) solution of $H_2O:NH_3(25\%):H_2O_2$ (30%) at 70ºC for 10 min followed by rinsing with Milli-Q water and drying in a stream of nitrogen gas. Before mounting the crystals in the flow chamber they were treated with UV/ozone for 30 min.
High quality Ruby Muscovite mica grade V-4, was purchased from SPI Supplies, USA.

*SbpA bacterial cell surface layer protein (S-layer)* was isolated from *Lysinibacillus sphaericus* CCM 2177 (former *Bacillus sphaericus*). Growth in continuous culture, cell wall preparation, extraction of S-layer protein with 5M guanidine hydrochloride (GHCl), dyalization and further centrifugation were carried out according to literature procedure [32]. The SbpA monomer solution used for recrystallization experiments was adjusted with Milli-Q water to a concentration of 1 mg mL$^{-1}$. Tris buffer containing protein monomers (protein: buffer volume ratio of 1:9) was used for recrystallization experiments.

*Lipid vesicles preparation:* The lipids were dissolved in chloroform and a film of lipid molecules was formed after the evaporation of the organic solvent under nitrogen stream and dried under vacuum for more than 1h. The obtained film was hydrated with HEPES buffer under vortexing in order to accelerate lipids to come in suspension. The lipids were then assembling with the hydrophobic part inside, forming multilamellar liposome vesicles (MLVs). In order to prepare large unilamellar vesicles (LUVs), MLV solutions were extruded several times through a polycarbonate membrane (100 nm diameter pores size) mounted in an extruder, at a temperature higher than the transition temperature (Tm) of the used lipid.

*Dynamic light scattering (DLS)* measurements were carried out with a Nano ZS ZEN 3600 Zetasizer from Malvern Instruments.

*Electrophoretical mobility* of the formed liposomes was measured with a Nano ZS ZEN 3600 Zetasizer from Malvern Instruments using Smolukowski approximation [33]. All measurements were carried out in HEPES solution.

*Quartz Crystal Microbalance with Dissipation Monitoring (QCM-D):* Adsorption and viscoelastic studies on S-proteins adsorbed on lipidic systems were carried out with a QE401 (electronic unit)/QFM401 (flow module) instrument from Q-sense AB (Gothenburg, Sweden) at 25ºC.

*Atomic Force Microscopy (AFM) measurements* were performed in aqueous solution (0.1 M NaCl), operating at room temperature in tapping mode (scan rate 1 Hz) with a Nanoscope V multimode (Veeco Instruments, Santa Barbara, CA). Silicon nitride ($Si_3N_4$) cantilevers with nominal spring constant of 0.1 N m$^{-1}$ were used.

*Nuclear Magnetic Resonance (NMR) studies* were performed with a Bruker Avance 500MHz spectrometer equipped with a 5 mm double resonance inverse probe. One-pulse experiments were recorded with 15 seconds of recycled delay and 64 transients. The proton spectral width of 8000 Hz and a total of 64k points were used with 90-degree pulse of 7.5 µs. The data was zero-filled to 128k points and then it was Fourier transformed. All the spectra were processed with Bruker TOPSPIN software.

*Langmuir-Blodgett (LB) film preparation*: Lipid monolayers were prepared with a Langmuir trough (Riegler & Kirstein Berlin). Lipids dissolved in chloroform in total concentration 0.25 mg/mL were spread onto Tris buffer subphase and left for solvent evaporation for 10 min. The lipid monolayer was compressed using a constant barrier speed of 12 Å molecule$^{-1}$ min$^{-1}$ to a pressure at which the monolayer does not reach the collapse. SbpA protein solution (1 mg mL$^{-1}$) was injected under the lipid monolayer and left overnight for protein adsortion (recrystallization).

*Transmission Electron Microscopy (TEM) studies* were performed with a JEM-2100F /UHR Pole Piece (JEOL, Japan) microscope with a 2k x 2k U-1000 CCD camera (GATAN, UK). The S-layer/lipid system prepared in the Langmuir-Blodgett trough was transferred onto formvar/carbon electron microscope grids. The grids were carefully placed onto the interface and removed after several seconds and then protein films were chemically cross-linked with 0.5% glutaraldehyde solution (in 0.1 M potassium phosphate buffer, pH 7.2) for 15 min and negatively stained with 0.1 % uranyl acetate in water for 15 min [34].

**Results and Discussion**

*The interaction of DOPC with SbpA protein*

The first studied case was the interaction of SbpA protein with the zwitterionic unsaturated 1,2-dioleoyl-sn-glycero-3-phosphocholine (DOPC). It is known that this lipid is able to form a fluid lipid bilayer when adsorbs on SiO$_2$ substrate [35]. DOPC vesicles were prepared as described in *Material and Methods* section, and dynamic light scattering and electrical mobility measurements characterized their size and apparent charge, respectively. It was found that the vesicle size was in the range of (148.0±1.3) nm; while the zeta potential, calculated from the electrophoretic mobility using the Smolukowski approximation [33] took a value of about (0.20±0.06) mV.

The purity of DOPC lipid and its ability to form vesicles was studied by NMR spectroscopy. Figure 2 presents $^1$H NMR spectrum of pure DOPC lipid in chloroform. The spectrum is free of any impurities and shows signal arising from –CH$_3$ groups marked by red bullets near to 1 ppm and also signal from -CH$_2$- groups marked by blue bullets between 1-1.5 ppm. Figure 6.2 presents also signals corresponding to –CH$_3$ groups (red color) and -CH$_2$- groups (blue color). The increase in line width corresponds to an increase in size of the molecules. This can be due to the aggregation of lipid molecules. Also, the decrease in the signals is due to the lack of molecules mobility. In addition, DOSY (diffusion order spectroscopy) experiments showed that the 2 signals come from a bigger molecule than from a single lipid molecule (data not shown).

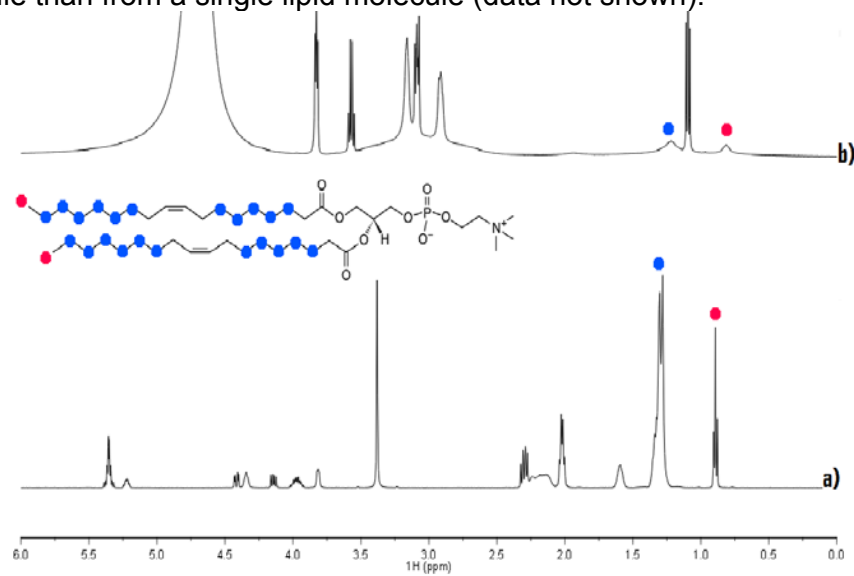

Figure 2. The $^1$H NMR spectrum of DOPC lipid in chloroform (a) and DOPC vesicles in HEPES buffer (b).

The adsorption of DOPC vesicles on silicon crystals was monitored using QCM-D as can be seen in Figure 3. At t = 1230s, DOPC vesicle solution (0.1 mg/mL) was injected. Vesicle adsorption induced a decrease in frequency (f) of 31 Hz. Vesicle rupture took place after approximately 50 s causing a final change in frequency up to 23 Hz, which is typical for lipid bilayer formation. Simultaneously to the frequency variation, the change in the dissipation (D) was recorded.

When the vesicles were adsorbed, the dissipation increased to 4 $\times 10^{-6}$, decreasing to about 0.66 $\times 10^{-6}$ after bilayer formation. This shows that the formed lipid bilayer was more rigid than the lipid vesicles. At t = 2050s, HEPES buffer was injected and f and D remained stable upon rinsing, indicating no vesicle desorption. Further, the next experimental step was the adsorption of S-layer protein on a DOPC bilayer. At t = 2680s, S-protein solution was introduced and left for adsorption during 30 min. After rinsing, an insignificant decrease in frequency (1 Hz) took place; while the dissipation increased with 1 $\times 10^{-6}$.

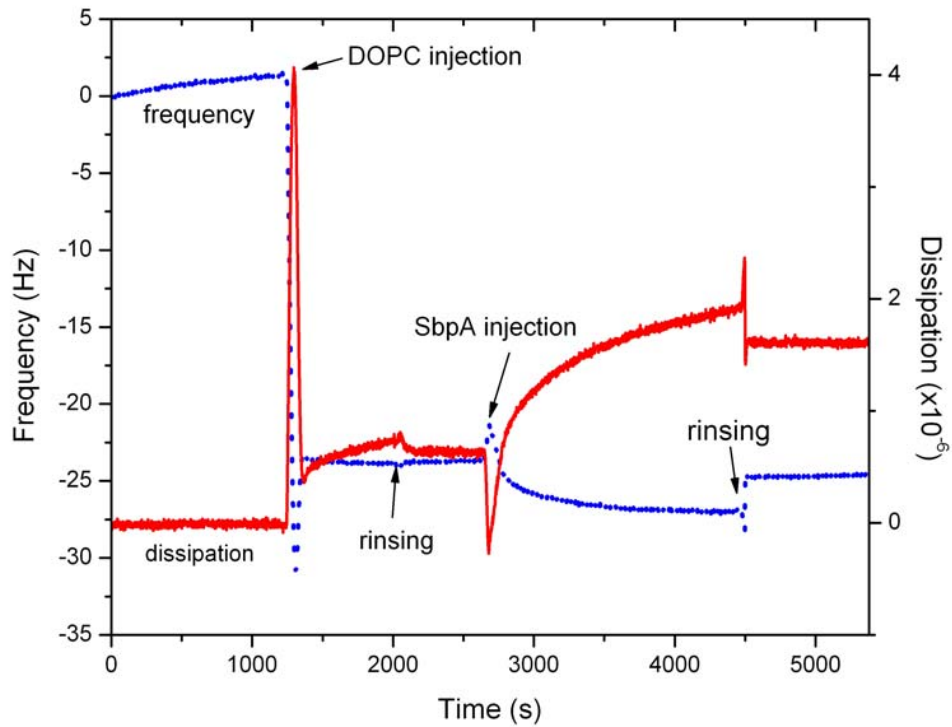

Figure 3. Changes in frequency (blue line) and dissipation (red line) for DOPC vesicle adsorption and S-layer protein adsorption as a function of time (the 5[th] overtone is shown). The measurement indicates bilayer formation after vesicles rupture, but no significant S-protein adsorption. A possible explanation could be that the DOPC lipid bilayer is to fluid (soft) at 25ºC presenting no available binding sites for protein deposition. From other studies, it is known that SbpA will absorb on "hard" surfaces, which confer little or none translational freedom of the molecules responsible for protein binding, such as positive liposomes [27], polyelectrolyte multilayers [36] or mica [38].

The QCM samples were characterized with AFM. Figure 4 shows a height AFM image of QCM sample (a), while the recrystallization of SbpA protein on silicon wafers (this can be considered as control experiment) is presented in (b). As can be seen, SbpA recrystallization is characterized by p4 crystalline structure [36] as shown by Fast Fourier Transform (FFT, placed at the bottom right corner). On the contrary, sample from QCM experiment does not show protein recrystallization.

lipid-bacterial protein interaction

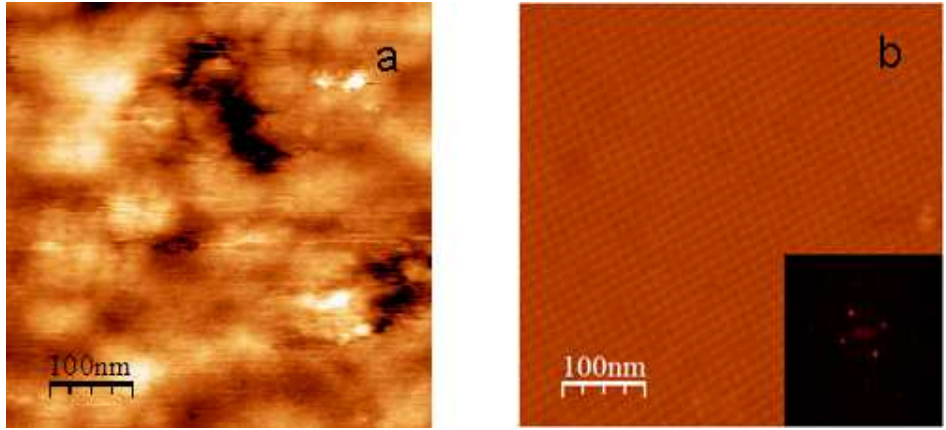

Figure 4. (a) Height AFM image of QCM sample (b) height AFM image of SbpA protein adsorbed on silicon and the corresponding FFT (at the right corner). The vertical scale of both images is 15 nm.

Since silicon crystals did not provide satisfactory results, another support (mica) was used. Mica is a well-known substrate for lipids and S-layer adsorptions [37, 38].
In a similar way to the QCM experiment, bilayer formation and S-protein adsorption were monitored by AFM as a function of time (see Figure 5). DOPC vesicle solution (0.1 mg/mL) was injected and left for adsorption for 30 min. Afterwards, the topography of the sample was imaged (a). The roughness of the system increased slightly, being 0.36 nm. Surface profile analysis (see green line in (b)) shows that the difference in thickness between the mica and the adsorbed lipid is about 5 nm, which should correspond to a lipid bilayer. After proving the lipid bilayer formation, the last step was to introduce the bacterial protein to the system. SbpA protein was injected and left for adsorption for 1h. The AFM picture shows that the typical nanostrtucture corresponding to S-layer formation is present all over the surface (c). Thus, recrystallization of SbpA protein on top of DOPC bilayer formed on mica was successful. The roughness of the SbpA protein adsorbed on mica covered with DOPC was 0.57 nm, higher than for mica and mica covered with DOPC. Surface profile analysis (see green line in (c) shows that the difference in thickness between the DOPC support and the adsorbed SbpA protein is about 8 nm, which should correspond to a protein monolayer.

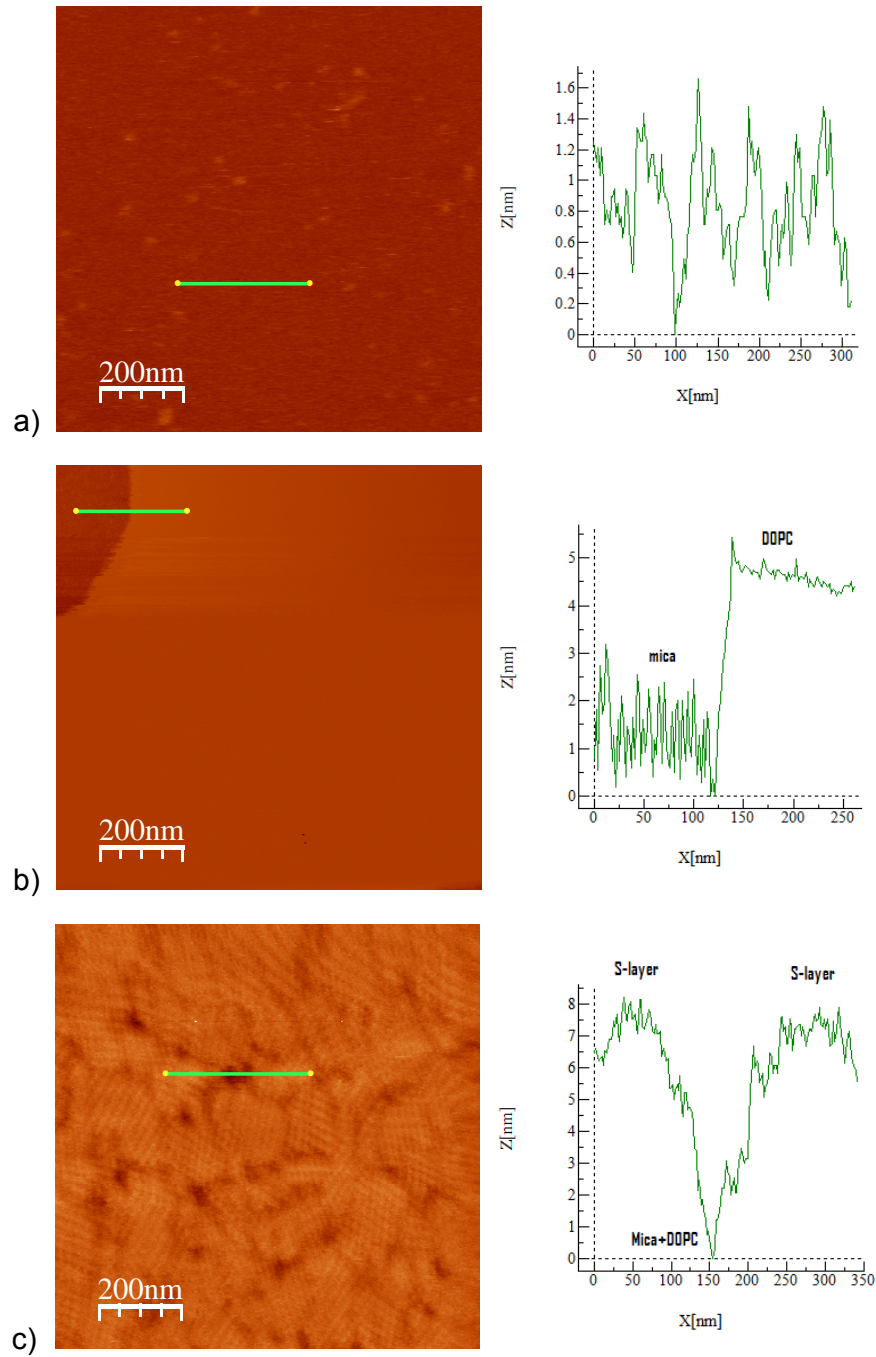

Figure 5. (a) Height AFM image and surface profile analysis (right side) of mica surface at t=0; (b) Height AFM image and surface profile analysis of mica covered with DOPC at t=30 min; (c) Height AFM image and surface profile analysis of SbpA protein adsorbed on DOPC bilayer at t=90 min. All images have a vertical scale of 15 nm.

Both silicon and mica supports allowed the formation of DOPC bilayer on top of them, but SbpA protein adsorption took place only on DOPC adsorbed on mica. Since silicon and mica are hydrophilic and negatively charged under our experimental conditions, the difference in surface chemistry and roughness (1.3 nm for silicon crystal) should influence lipid adsorption; and therefore the availability of the lipid substrate for the protein. Taken into account that the protein will need a system with less degree of freedom (lower lipid mobility) to form a stable layer, this result suggests that the lipid on mica is less mobile than on silicon.

*The interaction of DMPC with SbpA protein*

After using an unsaturated lipid, like DOPC, we used an zwitterionic saturated lipid: 1,2-dimyristoyl-sn-glycero-3-phosphocholine (DMPC). As can be seen, both lipids share the headgroup structure, which is of zwitterionic nature. As in the previous section, the size and the electrophoretic mobility of the vesicles were characterized. Freshly formed DMPC liposomes had size in the range of (106.0±0.5) nm, while the zeta potential took a value of (0.33±0.09) mV, which is in agreement with the zwitterionic nature of the lipid. It is important to mention that we used polyelectrolyte multilayers (PEM) as direct lipid support because the PEM-lipid system has been proposed as a biomimetic membrane in the last decade. In this way, we introduce (PSS) poly(sodium 4-styrenesulfonate) in the system, a negative surface with a more hydrophobic behavior than mica and silicon.

The coupling of lipid molecules to polymer components in a planar biomimetic model membrane made of DMPC lipid bilayer supported by polyelectrolyte multilayers was studied by neutron reflectometry [39].

In this work, we have monitored the adsorption of DMPC (0.1 mg/mL) on polyelectrolyte multilayer (PEM) and further SbpA protein adsorption by QCM-D (see Figure 6). At t = 800s, the silicon surface is exposed to the Poly(ethylenimine) (PEI) solution, resulting in a decrease in frequency (f) and an increase in the dissipation (D). At t = 847s, polyelectrolyte deposition is interrupted by exchange to Milli-Q water. At t = 1617s, Milli-Q water is changed to Poly(sodium 4-styrenesulfonate) (PSS) solution. The polyelectrolyte multilayer deposition continued with the injection of Poly(allylamine hydrochloride) (PAH) solution at t = 2340s, PSS solution at t = 3100s, PAH at t = 3780s and again PSS at t = 4500s. A uniform decrease in the frequency is noticed after every polyelectrolyte layer deposition. At t = 5140s, DMPC solution was injected and left for adsorption and then was changed to HEPES buffer at t = 9895s. DMPC vesicles adsorb and remain intact on PSS-terminated PEM as can be deduced from the decrease in the frequency (158 Hz) and the increase in dissipation of about $30 \times 10^{-6}$. According to the Sauerbrey equation [40], the decrease in the frequency corresponds to a surface mass of about 2797 ng/cm$^2$ due to DMPC vesicle adsorption.

Finally, S-layer protein solution was injected in the experimental cell at t = 11762s and was left for adsorption during one hour. No remarkable SbpA protein adsorption took place; the decrease in frequency of about 23 Hz corresponds to a surface mass of 407 ng/cm$^2$. In a previous work, it has been reported that SbpA protein layer formation induces a change in frequency of about 92 Hz [41], therefore we can conclude that SbpA protein adsorbs randomly on DMPC vesicles without forming a protein monolayer or bilayer. The adsorption of S-layer proteins also produces an increase in the dissipation (ca.

5 x10$^{-6}$), which indicates that the final hybrid-protein layer is more viscoelastic than DMPC vesicles.

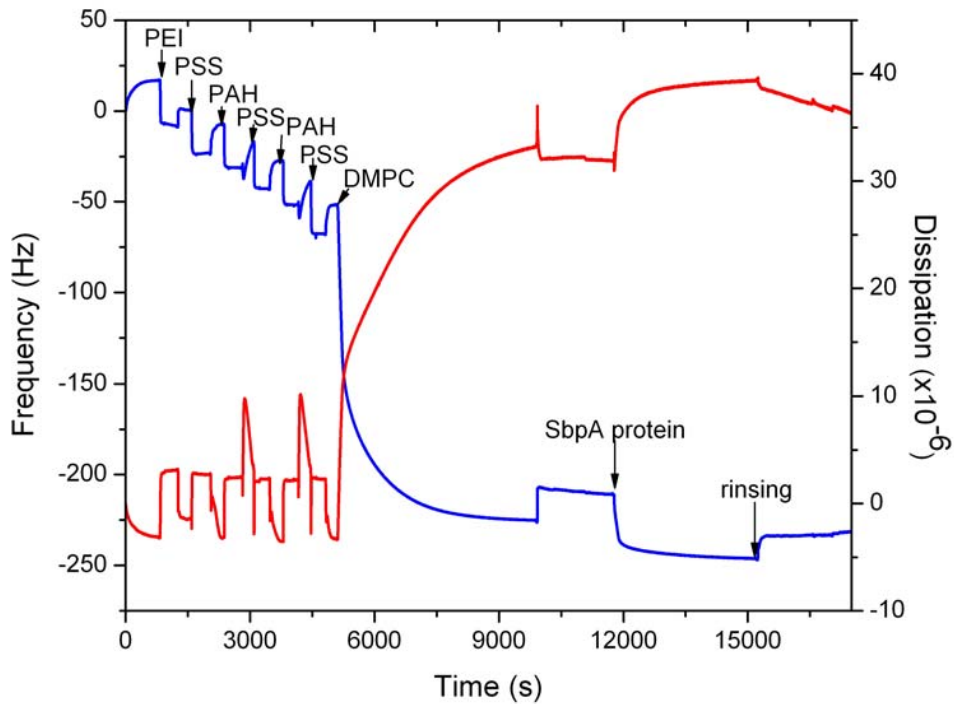

Figure 6. Changes in frequency (blue line) and dissipation (red line) for DMPC vesicles adsorption on PEM, and further S-layer protein adsorption as a function of time. Note the monotonous decrease in frequency due to PEM formation followed by the strong decrease in frequency caused by DMPC vesicles adsorption (158 Hz) and the slight decrease in frequency (23 Hz) due to SbpA protein adsorption. The strongest change in dissipation is produced by the lipid-protein system. The 5$^{th}$ overtone is shown here.

*The interaction of DPPC with SbpA protein*

The difference between DMPC and 1,2-dipalmitoyl-sn-glycero-3-phosphocholine (DPPC) is the alkyl chain length. DPPC has two more -CH$_2$- groups, which imply a higher melting temperature (41ºC) than DMPC (23ºC). When we performed the DMPC experiments at room temperature, the lipids were in a fluid phase (disordered chains), while in the experiments performed in this section, DPPC was in a gel phase. In this way, we were able to check if the thermodynamic state of the lipid influenced the lipid-protein interaction since the lipid headgroup remained the same.

The size and the apparent charge of the freshly formed DPPC liposomes were measured by DLS and electrophoretic mobility. DPPC liposomes presented a size of (155.0±2.1) nm and the zeta potential was close to zero (0.8±0.03) mV.

The adsorption of DPPC vesicles on silicon crystals was monitored using QCM-D as can be seen in Figure 7. At t = 920s, a DPPC vesicle solution (0.1 mg/mL) was injected. A decrease in frequency of 310 Hz was recorded; this can be interpreted as DPPC vesicle adsorption on silicon, remaining intact with time (no fusion). This value is higher than the one obtained for DMPC vesicle adsorption on PSS-terminated PEM. This is not unexpected since DPPC vesicles are larger than DMPC vesicles, and the DPPC molecule is heavier than DMPC. Vesicle adsorption is correlated with an increase in dissipation, which took a value of 20 $\times 10^{-6}$. At t = 1990s, HEPES buffer was injected and f and D remain quite stable upon rinsing indicating that the adsorbed vesicles are stable. Further, we introduced S-layer protein solution in the system at t = 3475s. The system equilibrated for 1h. Protein adsorption induced a decrease in frequency of 52 Hz, and an increase in dissipation of about 15 $\times 10^{-6}$. However, after the rinsing step, the frequency and the dissipation reached similar values as for DPPC vesicles adsorption.

This measurement is a proof that SbpA protein is removed after rinsing step and therefore, does not have strong affinity to DPPC.

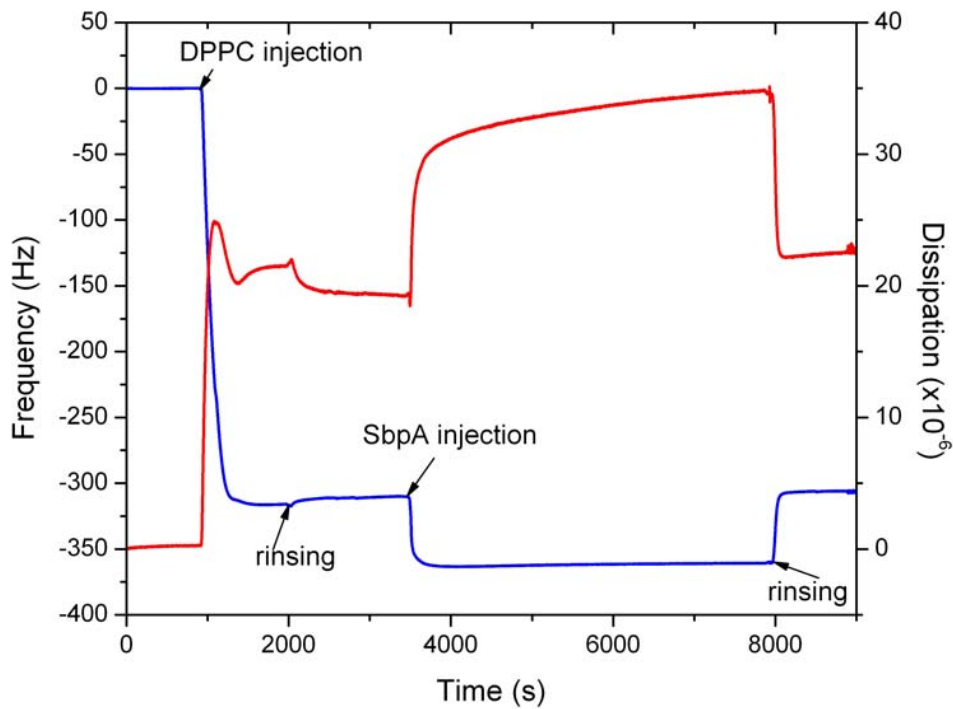

Figure 7. Changes in frequency (blue line) and dissipation (red line) for DPPC vesicles adsorption and further S-layer protein adsorption as a function of time. From the measurement it can be clearly stated that DPPC vesicles adsorb strongly on silicon crystals. It can be also observed that SbpA protein desorption occurred during washing step. The 5$^{th}$ overtone is shown here.

In order to clarify the absence of protein on DPPC vesicles, AFM experiments were carried out. Figure 8 shows that DPPC vesicles (see white empty circles) of diameter 200 nm were adsorbed on silicon crystals.

The adsorption of DPPC vesicles (200 nm diameter) on silicon produces a deformation of the vesicles. This could be deduced from surface profile analysis (see green line), which shows a difference in height between the vesicles and the silicon substrate between 50-80 nm. No crystalline S-protein structure could be observed on the vesicle surface.

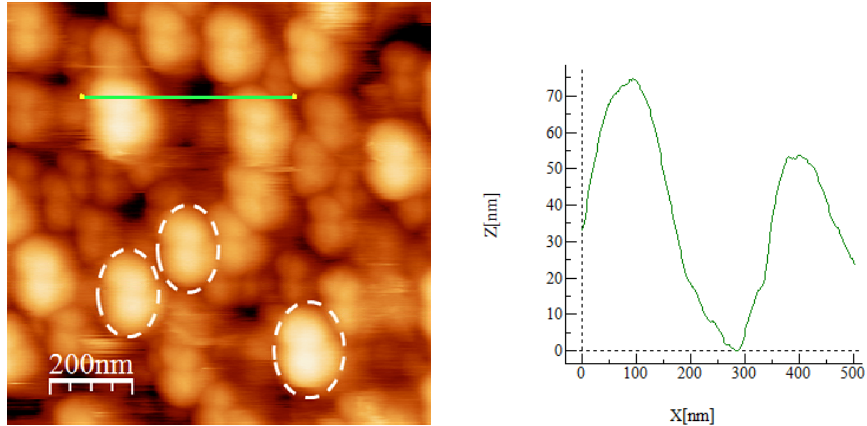

Figure 8. On the left: height AFM image of DPPC vesicles (white circles) adsorbed on silicon. Neither S-protein, nor typical S-layer structure could be observed. The vertical scale of the image is 77 nm. On the right: a profile analysis (along the green line) shows that the height of the vesicles is from 50 to 70 nm.

This AFM result is in agreement with QCM-D results. Both experimental techniques confirmed that SbpA protein has no affinity for DPPC vesicles. In this case, we have proved that the variation of the thermodynamic state of the alkyl lipid chain maintaining the same headgroup (PC) does not lead either to SbpA protein adsorption or to protein crystal formation.

*The interaction of DMPG with SbpA protein*

Until now we have used uncharged lipids (due to zwitterionic nature of PC). In this section, we are using (DMPG) 1,2-dimyristoyl-sn-glycero-3-[phospho-rac-(1-glycerol)], a saturated lipid, with the same chain length as DMPC but different (charged) headgroup. DMPG as a saturated anionic phospholipid constitutes one of the major membrane phospholipids.
Freshly formed DMPG liposomes had a size in the range of (105.0±2.2) nm and were negatively charged (-26.3±1.8) mV as measured by dynamic light scattering and electrophoretic mobility. The presence of DMPG liposomes was also confirmed by TEM as is shown in Figure 9. Vesicles with size in the range 100-200 nm were observed (which is in agreement within error with DLS measurements).

lipid-bacterial protein interaction

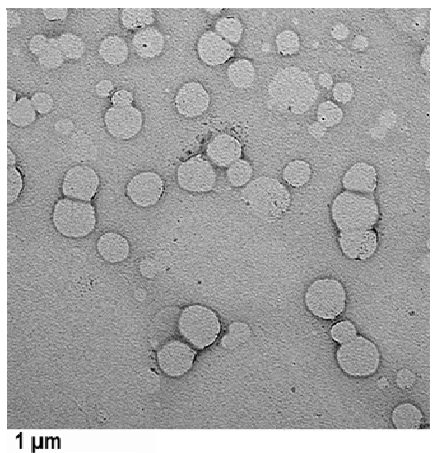

Figure 9. TEM image of DMPG vesicles deposited on carbon grids. The vesicles were negatively stained with uranyl acetate. The size range of the DMPG vesicles is in agreement with DLS measurements within error.

The purity of DMPG lipid and the presence of the vesicles have been studied by NMR spectroscopy. Figure 10 presents $^1$H NMR spectrum of DMPG lipid in chloroform.
The spectrum is free of any impurities shows signals arising from –CH$_3$ groups marked by red bullets near to 1 ppm and also signals from -CH$_2$- groups marked by blue bullets between 1-1.5 ppm.  The deuterated water present in chloroform gives the signal at 1.55 ppm. Figure 10 presents also signals corresponding to –CH$_3$ groups (red color) and -CH$_2$- groups (blue color). The increase in line width corresponds to an increase in size of the molecules. This can be due to the aggregation of lipid molecules. Also, the decrease in the signals is due to the lack of molecules mobility. In addition, DOSY (diffusion order spectroscopy) experiments showed that the 2 signals come from a bigger molecule than from a single lipid molecule.

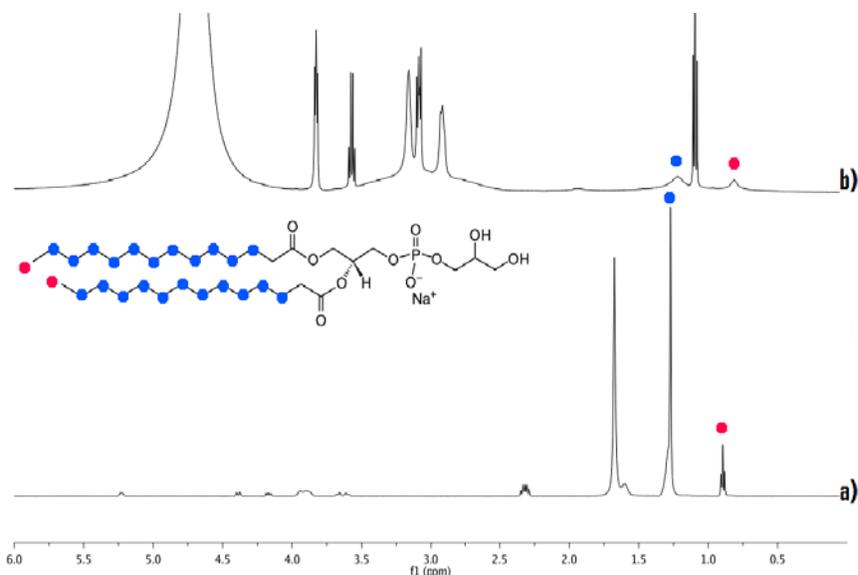

Figure 10. The $^1$H NMR spectrum of DMPG lipid in chloroform (a) and DMPG vesicles in HEPES buffer (b).

Vesicle adsorption on silicon crystals was monitored by QCM-D as can be seen in Figure 11. At t = 2000s, DMPG vesicles solution (0.1 mg/mL) was injected, causing a decrease in frequency of 20 Hz and an increase in dissipation up to 8 x$10^{-6}$. This change in frequency indicates a bilayer lipid formation at one step, a different mechanism from the DOPC case. At t = 5420s, HEPES buffer was injected and f and D remained quite stable upon rinsing indicating that the vesicles are adsorbed in a stable manner. Finally, S-layer protein was introduced in the system. At t = 6320s, S-layer protein solution was injected and left for adsorption during more than 1h. It can be clearly seen in Figure 6.11 that SbpA protein adsorption induces a change in frequency of 93 Hz which, applying Sauerbrey equation corresponds to a surface mass of 1646 ng/cm$^2$. This value is in agreement with results obtained for SbpA adsorption on PEM and silanes [41, 42].

The adsorption of SbpA protein increased slightly the dissipation (2 x$10^{-6}$), conferring the lipid-protein system a higher viscoelasticity than the lipid layer itself.

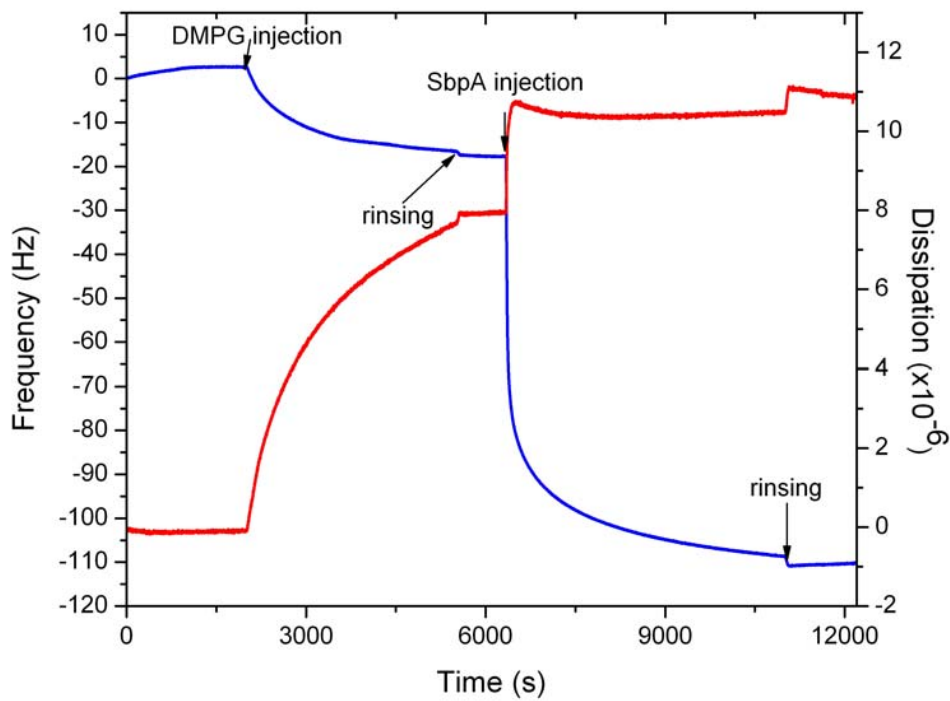

Figure 11. Changes in frequency (blue line) and dissipation (red line) for DMPG vesicles adsorption, and further S-layer protein adsorption as a function of time. A decrease in frequency of 20 Hz is related to DMPG bilayer formation. More remarkable is the change in frequency (93 Hz) after SbpA protein injection, which means a very strong protein-lipid interaction. The 5$^{th}$ overtone is shown here.

After adsorbing bacterial protein on a DMPG bilayer we proceed to study the SbpA interaction with DMPG vesicles. Mica was the appropriate support for DMPG vesicle adsorption. A lipid solution of 0.1 mg/mL was adsorbed on mica during 30 min. Afterwards, SbpA protein was injected into the system, which equilibrated for 1h.

Height AFM measurements give an overview image of the system SbpA-covered vesicles, the diameter of the vesicles being in the range of 100-200 nm (a). However, from this picture it is not possible to distinguish if the protein is covering the vesicles. In order to elucidate this question, we made a zoom of (a). Height AFM image (b) shows the presence of crystalline structure of SbpA protein on DMPG vesicles.

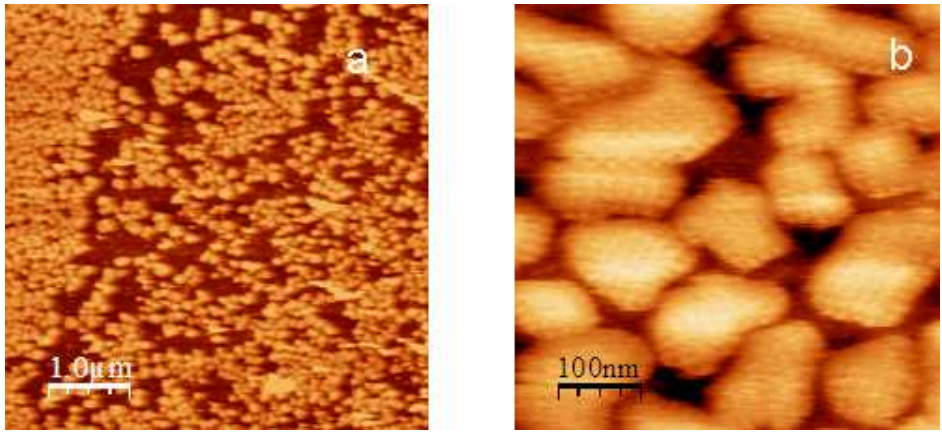

Figure 12. Height AFM images of: (a) DMPG vesicles possibly coated with SbpA protein; (b) a zoom of picture a shows recrystallized SbpA protein on DMPG vesicles. The vertical scale of the images is 20 nm.

*The interaction of DOPC/DOPS (4:1) mixture with SbpA protein*

In the last sections, we studied the interaction between protein and lipid layers composed of a single type of phospholipid. However, this is not the normal situation in biological membranes, since they are composed of lipid mixtures, proteins, etc.

In this section, we studied the SbpA protein interaction with phospholipid mixtures. These mixtures allow us to vary not only the exposed lipid surface charge to the protein, but also the lipid headgroup size and the fluidity of the lipid layer. The first used lipid mixture consisted of DOPC (1,2-dioleoyl-sn-glycero-3-phosphocholine) and DOPS (1,2-Dioleoyl-*sn*-Glycero-3-[Phospho-L-Serine]) in a 4:1 molar ratio. It has to be said that both lipids are unsaturated, DOPC being uncharged, while DOPS has negative charge due to serine group (see Figure 1). Another reason to use this lipid mixture is that it forms a lipid bilayer on silicon or mica substrates [43]. The size and the apparent charge of DOPC/DOPS (4:1) liposomes were characterized by DLS and electrophoretic mobility. Liposomes had size in the range of (130±3) nm and the zeta potential was (-12.0±0.3) mV, which is slightly negatively charged as expected from DOPS molecules.

lipid-bacterial protein interaction

The adsorption of DOPC/DOPS (4:1) vesicle mixture on silicon was monitored using QCM-D as can be seen in Figure 13. At t = 765s, vesicle mixture solution (0.1 mg/mL) was injected. A decrease in frequency of 70 Hz and an increase in dissipation of $8 \times 10^{-6}$ were recorded as a consequence of vesicle adsorption. Vesicle rupture led to bilayer formation as the final change in frequency shows (26 Hz), while the dissipation reached a value close to zero. Finally, at t = 4240s, SbpA protein solution was injected. The system was left 1h for equilibration. After rinsing, the final decrease in frequency of about 72 Hz and increase in dissipation of about $14 \times 10^{-6}$ was registered. The change in frequency after SbpA protein adsorption corresponds to a surface mass of 1274ng/cm$^2$, which is a bit lower than a S-protein bilayer and slightly larger than a protein monolayer. This measurement shows that the addition of negative lipid is enough to force SbpA protein adsorption on the (4:1) DOPC/DOPS lipid mixture bilayer comparing with the DOPC bilayer.

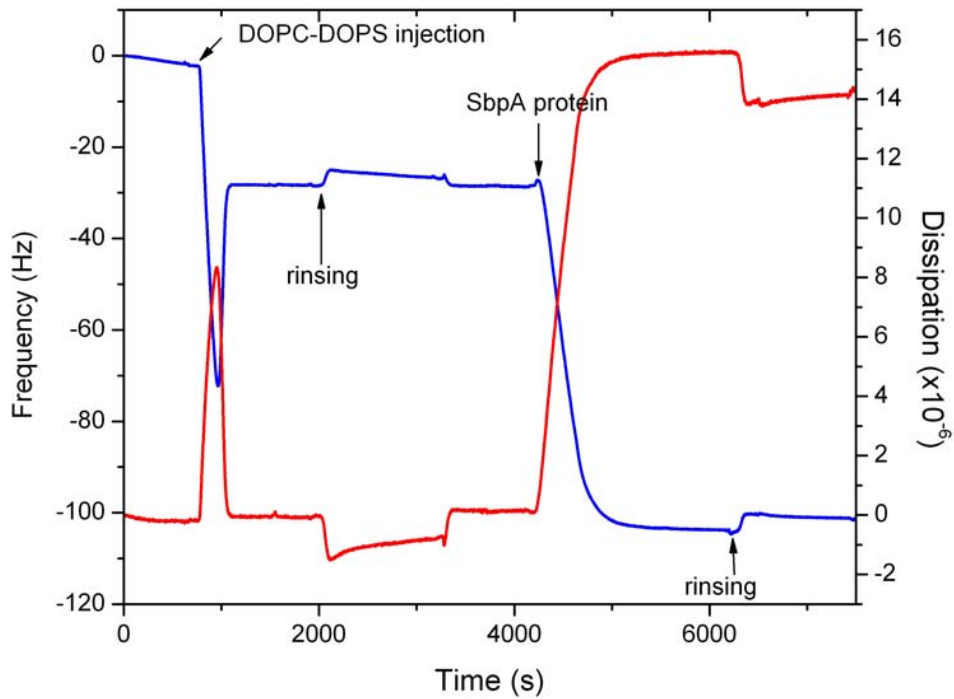

Figure 13. Changes in frequency (blue line) and dissipation (red line) for DOPC/DOPS (4:1) vesicles mixture adsorption and further S-layer protein adsorption as a function of time. Bilayer formation occurs after vesicle rupture. SbpA protein adsorbs on the formed bilayer. A charge ratio of DOPC/DOPS (4:1) is enough to adsorb SbpA protein. The 5$^{th}$ overtone is shown here.

Structural information of the protein/lipid mixture system was studied by AFM. Figure 14 shows a height AFM image and the corresponding surface profile analysis of SbpA protein adsorbed on DOPC/DOPS mixture. It can be seen that protein adsorption did not lead to a regular crystalline layer (Fast Fourier Transform at the top right corner shows no regularity). The measurement shows a non-homogeneous surface with a roughness of 1.5 nm, likely composed of protein-lipid mixture. Regular spots (white empty circles) of about 50 nm in diameter can be observed. Surface profile analysis (see green line) shows that the largest thickness is about 4-5 nm (the thickness of a lipid bilayer), while the length of the aggregates is around 50-100 nm.

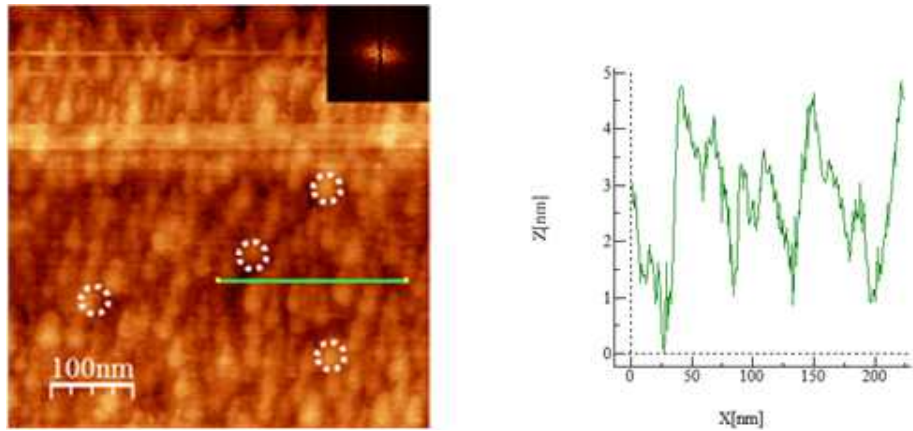

Figure 14. Left: Height AFM image of SbpA protein adsorbed on DOPC/DOPS (4:1) mixture and the corresponding FFT at the top right corner, showing no crystalline protein structure. The vertical scale of the image is 20 nm. Right: Surface profile analysis (along the green line) shows a thickness difference of about 4-5 nm, and aggregates around 50-100 nm in length.

Although a DOPC/DOPS (4:1) mixture and DOPC form a lipid bilayer on silicon, only the mixture is able to attract SbpA proteins without obtaining a crystalline protein layer.
SbpA protein adsorption can be due to the negative nature of the lipid bilayer since the viscoelastic properties of DOPC bilayer and the lipid mixture bilayer are similar as shown by QCM-D (see Figure 3 and Figure 13).

*The interaction of DOPC/DMPG (1:1) mixture with SbpA protein*

The introduction of negative charge in a lipid mixture composed of unsaturated DOPC and DOPS led to the adsorption of SbpA protein, although the viscoelastic properties of both type of bilayers were similar.
This lipid mixture should change simultaneously the surface charge and the rigidity of the lipid layer. The mixture is composed of DOPC (1,2-dioleoyl-sn-glycero-3-phosphocholine) and DMPG (1,2-Dioleoyl-*sn*-Glycero-3-[Phospho-L-Serine]) in a molar ratio of 1:1. Previously it has been shown that DOPC vesicles led to a lipid bilayer formation, while DMPG vesicles attracted SbpA proteins.
The size and the apparent charge of DOPC/DMPG (1:1) liposomes were characterized by DLS and electrophoretic mobility. Liposomes had a size in the range of (186±3) nm and a zeta potential of (-16.0±0.3) mV. This value is very similar to the reported zeta potential of DOPC/DOPS mixture.
The adsorption of the DOPC/DMPG vesicles on silicon was monitored by QCM-D as can be seen in Figure 15. At t = 740s, vesicle mixture solution (0.1 mg/mL) was injected. A decrease in frequency of 66 Hz and an increase in dissipation of 5 x$10^{-6}$ indicate vesicle adsorption. At that point, the system remained in an intermediate state for approximately 1500s until vesicle rupture occurred leading to a change in frequency of 38 Hz, while the dissipation reached a value of (3 x$10^{-6}$). The usual values for bilayer formation are around 25 Hz and 1 x$10^{-6}$ for the change in frequency and dissipation respectively. Finally, at t = 3320s, SbpA protein solution was injected. The system was left for equilibration during 1h. No significant change in frequency was noticed, meaning that no significant SbpA adsorption on a (1:1) DOPC/DMPG mixture took place.

Although, it has been shown that DMPG vesicles attract SbpA proteins, the ability of DMPG to attract SbpA protein is lost when DMPG is mixed with DOPC in a lipid bilayer.

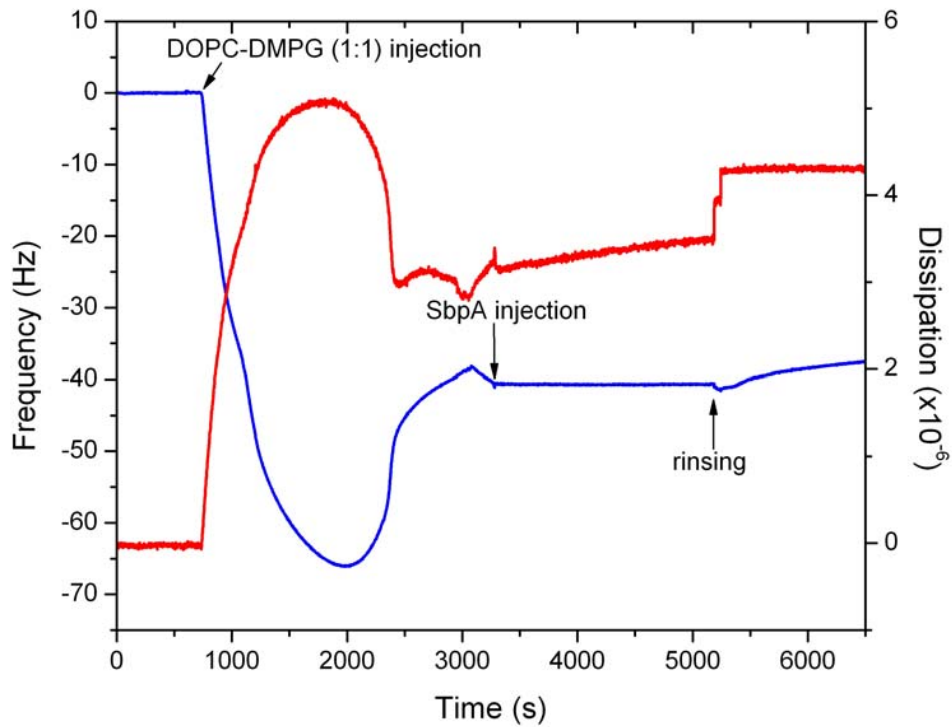

Figure 15. Representative measurement of the changes in frequency (blue line) and dissipation (red line) for DOPC-DMPG (1:1) vesicle adsorption. At 2000s, a minimum in the frequency is observed. However, the shape of the minimum is wider than the minimum observed for DOPC vesicle adsorption and DOPC-DOPS (4:1) vesicle adsorption. This should be connected to the kinetic behavior of vesicle rupture that in this case is slower. This slow kinetics ends with the formation of an intermediate state composed of vesicles and a lipid bilayer. The frequency and dissipation show that SbpA protein does not adsorb on a DOPC-DMPG (1:1) lipid mixture. The $5^{th}$ overtone is shown here.

Structural studies of SbpA protein adsorbed on a DOPC/DMPG (1:1) mixture were investigated by AFM (see Figure 16). Since the mixture DOPC/DMPG on silicon did not lead to SbpA adsorption, another support (mica) was used for vesicle adsorption. DOPC/DMPG (1:1) lipid suspension (0.1 mg/mL) was left to adsorb on mica surface for 30 min. Afterwards, SbpA protein (0.1 mg/mL) was introduced in the system, which was left 1h for equilibration. Figure 16 shows the topography of the protein-lipid mixture. Two different structures can be observed: the first one consists of patches with the regular protein structure (roughness of 0.6 nm), and the second one, consisting of very flat lipid domains (roughness of 0.3 nm) (bright areas). This argument is supported by surface profile analysis (see the green line in Figure 6.16), which shows a difference in thickness between the two structures of about 3-3.5 nm, a value that can be attributed to the thickness of a lipid bilayer.

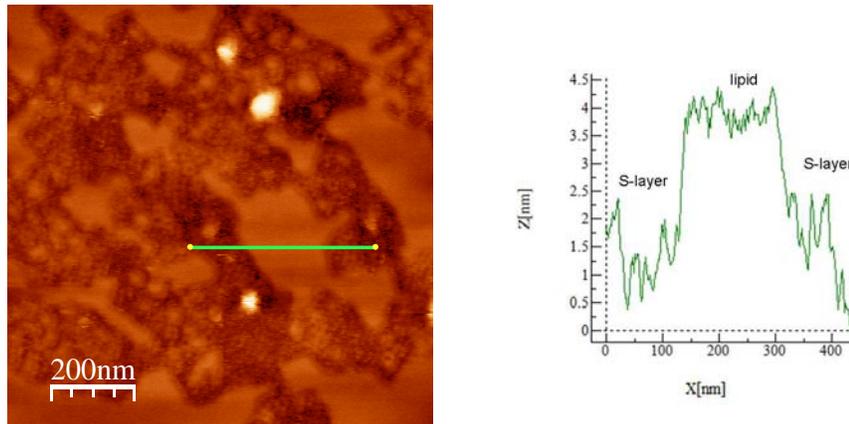

Figure 16. Left: Height AFM image and surface profile analysis of the SbpA protein/(DOPC/DMPG) lipid system on mica. Note the hybrid nature of the system, consisting of adsorbed protein and flat lipid domains, which are likely adsorbed on top of the protein layers. This result is an unexpected phenomenon, since indicates that the protein removed the adsorbed lipid layer, and opens a new question regarding the adsorption kinetics (competition) of SbpA protein and DOPC/DMPG (1:1) on mica substrate. Right: The cross section (along the green line) shows a height of about 3.5 nm on the S-protein. The vertical scale of the AFM image is 15 nm.

Further characterization of the protein/lipid mixture surfaces was carried out performing force-distance curves measurements different areas. Figure 17 shows a force-distance curve carried in the bright areas. The approaching curve shows a slight repulsion regime of about 5 nm range, before touching the surface. The load increases until a kink of about 5 nm appears. This means that the AFM tip has indented the lipid layer [44]. The retracting curve shows a small adhesion peak, which is related to the interaction between the AFM tip and the lipid molecules, reaching the zero force level at approximately 10 nm distance. Note that the picture presents an offset for the deflection error (zero force level) of -24 nm (the force is the deflection of the cantilever multiplied by its spring constant). Figure 18 shows a force-distance curve carried out in the protein areas. The approaching and retracting curves coincide; this is typical for S-layer structure [36]. The approaching curve shows that no indentation event takes place for loads four times higher than the previous case.

lipid-bacterial protein interaction

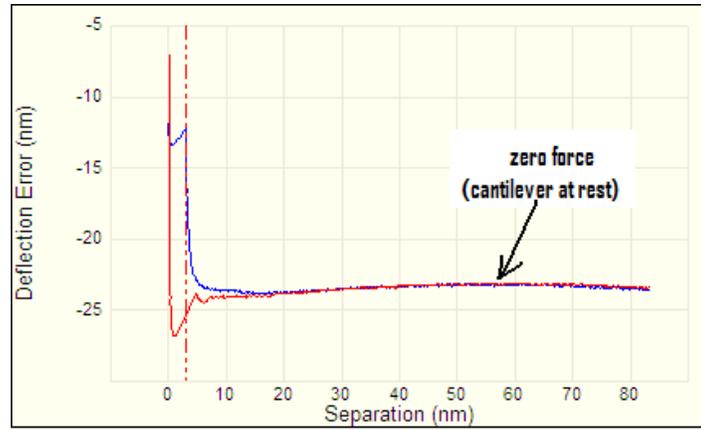

Figure 17. Force-distance curve taken on bright areas of Figure 16. The kink observed in the approaching curve is a measure of a bilayer thickness (5nm)

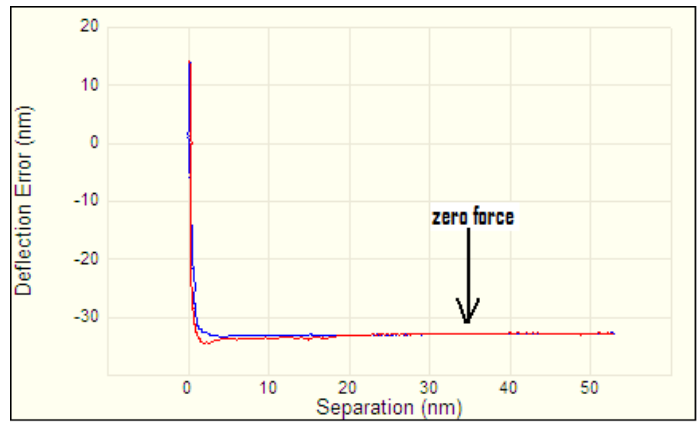

Figure 18. Force-distance curve taken on the regular protein structure of Figure 14. In this case, no indentation event is observed. The overlapping of approaching and retracting curves shows that the material is not viscoelastic (no hysteresis occurs). Furthermore, no protein unfolding events are observed.

The surface profile analysis and the force-distance measurements indicates the existence of a lipid bilayer on top of the S-layer structure. A hypothesis that should be examined in detail is that SbpA proteins remove the lipid molecules during the protein self-assembly process on the mica surface.

*The interaction of lipid monolayers with SbpA protein*

In the previous sections, bilayer formation or vesicle adsorption was led by the interaction between the lipid molecules and the substrate. However, we could not control either the thermodynamic state of the lipid, or the amount of the adsorbed lipid. In this section, we will work with lipid monolayers on a Langmuir trough. This allows to control the lipid phase behavior and the amount of the lipid present at the air/water interface. An advantage of this technique is that one can change and control the thermodynamic state of the lipid at air/water interface by compressing the layer with two barriers. Lipid monolayers were obtained after spreading the lipids at the air/water interface when the two barriers were far away from each other (maximum available surface). At this moment, the lipids are in a gas phase (there are no collisions between lipid molecules). By moving the barriers, the lipid molecules are forced to occupy a smaller available area. In this way, there is a moment when the lipid molecules start to interact with each other, and a surface pressure is measured. When this type of experiment is carried out at constant temperature, it is called isotherm. From former experiments it is known that SbpA protein adsorbs on solid supports like silicon wafers, mica, and "hard" well-organized surfaces, such as lipid monolayers, disulphides or silanes [42, 45]. Therefore, we prepared three "hard" well-organized lipid layers.

In order to achieve this goal, surface pressure-area (π-A) isotherms were measured in a aqueous subphase (0.5 mM Tris-HCl, 10 mM $CaCl_2$, pH 9 buffer) at 25ºC, for the following lipid systems: DPPC, DPPC/DOPS (1:1) and DPPC/DMPG (1:1). Representative isotherms are shown in Figure 19. The first isotherm (a) corresponds to uncharged DPPC monolayer. This isotherm is very well known [46]. The DPPC monolayer isotherm presents a plateau region at a surface pressure of 8 mN/m between molecular areas of 100-150Å$^2$, reaching a more condensed state for molecular areas bellow 100 Å$^2$. This condensed phase means that the lipid is well packed being a suitable candidate support for SbpA protein adsorption. Therefore, the chosen surface pressure for SbpA protein to adsorb was 30 mN/m (the lipid molecules are well-packed).

The second isotherm (b) corresponds to the DPPC/DOPS (1:1) lipid mixture. The addition of charged DOPS eliminates the fluid-crystalline coexistence phase (plateau region). The surface pressure remains zero until a molecular area of about 135Å$^2$ is reached. At that molecular area, lipid molecules start to interact with each other; this can be seen in the monotonous increase in the surface pressure with molecular area decreasing. Since at 28 mN/m, for a molecular area of 60 Å$^2$, the lipid molecules are well-packed, this surface pressure was chosen to adsorb SbpA protein on the lipid monolayer.

The last isotherm (c) corresponds to a lipid mixture of DPPC/DMPG (1:1).

The addition of charged DMPG also removes the plateau region of the DPPC isotherm. The surface pressure remains zero until a molecular area of about 150 Å$^2$ is reached and increased monotonously until 16 mN/m, which is almost half of the surface pressure of the two previous isotherms. SbpA adsorption was carried out at this surface pressure.

a)

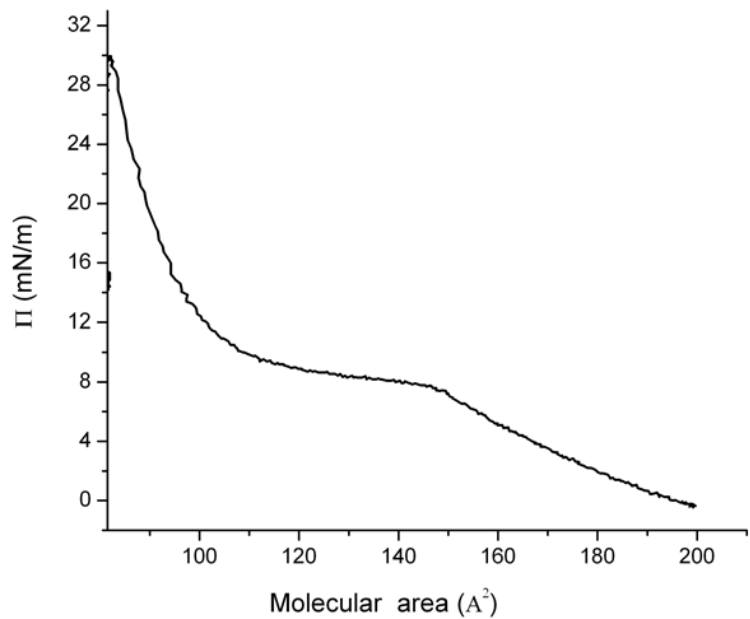

b)

lipid-bacterial protein interaction

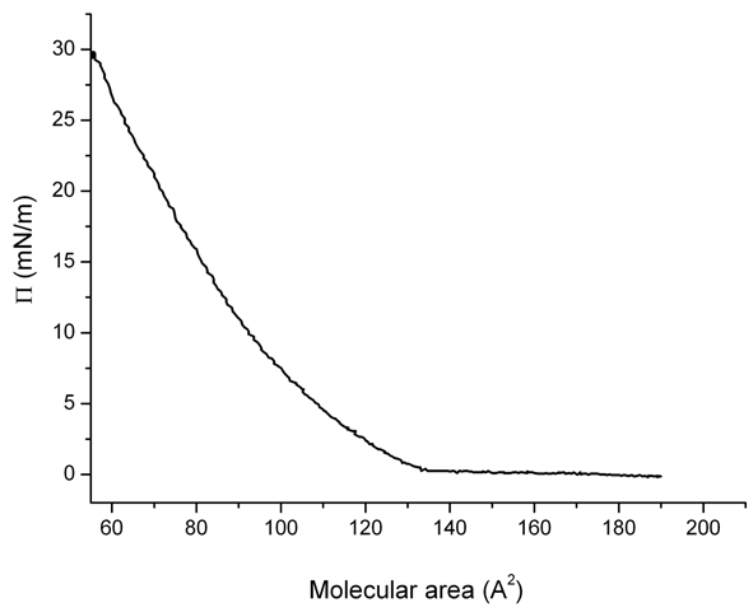

c)

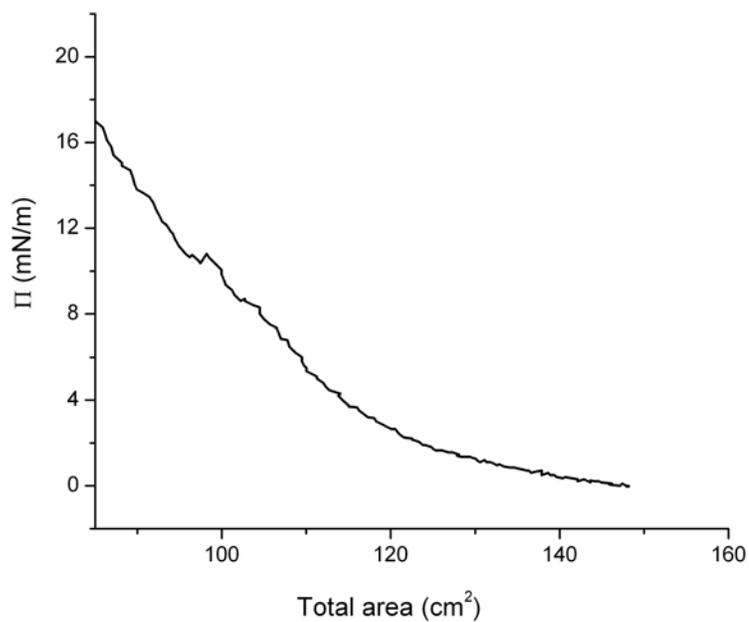

Figure 19. π-A isotherms of the following lipid systems: (a) DPPC, (b) DPPC/DOPS (1:1) and (c) DPPC/DMPG (1:1). Tris buffer was used as aqueous surface. The temperature was kept constant at 25ºC. Note that case (c) refers to total area.

After lipid monolayer formation, SbpA protein solution (3 mL) was injected under the lipid monolayer. Protein adsorption experiments were carried out overnight. Finally, the lipid/protein system was transferred to a formvar carbon grid. The structure of the protein/lipid system was investigated with TEM. EM-micrographs of SbpA protein adsorbed onto lipid monolayers are shown in Figure 20.

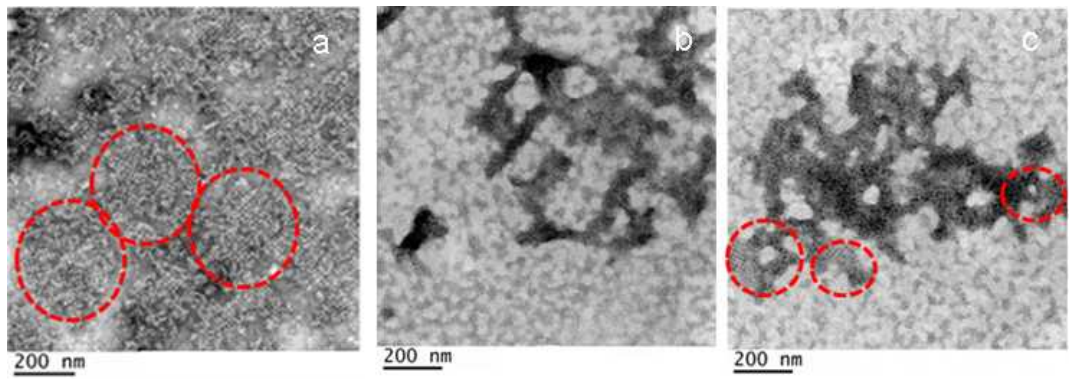

Figure 20. EM-micrographs of SbpA protein recrystallized onto lipid monolayers: (a) DPPC; (b) DPPC/DOPS (1:1); (c) DPPC/DMPG (1:1). Red circles show areas with recrystallized S-protein. Note: uranyl acetate makes protein look white.

A condensed DPPC monolayer is a suitable substrate for SbpA protein crystallization as several patches (red empty circles) with crystalline structure can be observed (a).

We have seen previously that SbpA protein does not adsorb on DPPC vesicles, while in this case, SbpA protein is able to form a protein layer due to the fact that the lipid monolayer is in a condensed state. Thus, closer interaction between the lipids and proteins occurs. The interaction is primarily electrostatic (probably dipolar) and the contact between the lipid monolayer and the adsorbed protein occurs through the "primary" binding sites on the protein surface [23]. Regular structures, which do not correspond to the known protein crystalline structure, are present in (b). Only small areas could be associated to a protein crystal. This experiment shows that a condensed monolayer made of DPPC/DOPS (1:1) mixture is not a good surface for S-layer crystallization but for S-layer protein adsorption. This fact is not unexpected since, before, we showed that anionic DOPS was responsible for S-protein adsorption when mixed with DOPC.

Some S-layer areas (red empty circles) can be observed on the DPPC/DMPG (1:1) mixture (c). It has to be said that DMPG vesicles are a good surface for SbpA protein recrystallization as shown previously. This process is not only driven by dipolar interactions (DPPC molecules) but also through point charge interactions (DMPG). In this section, we have shown that lipid monolayers can be suitable surface for SbpA recrystallization. However, many parameters have to be improved and investigated in the near future, such as the appropriate lipid ratio mixture, temperature conditions (the π-A isotherms) and solvent conditions (ionic strength, pH, etc); thus, many questions remain still unanswered.

**Conclusions**

The interaction of SbpA protein with saturated, unsaturated, charged and uncharged lipids has been investigated.
It has been found that zwitterionic DOPC bilayers, negatively charged DMPG vesicles, and DOPC/DMPG bilayers adsorbed on mica, together with DPPC and DPPC/DMPG monolayers, have high affinity for SbpA protein, leading to the formation of crystalline protein layers. Further investigations indicated a weak interaction between SbpA proteins and DOPC and DOPC/DOPS bilayers supported on silicon. Due to the nature of this interaction, no protein crystal could be formed. Finally, two lipid systems, DMPC vesicles adsorbed on polyelectrolytes multilayers and DPPC vesicles supported on silicon, did not have any affinity for SbpA proteins.

In this work we have shown the importance of the surface chemistry and roughness of the substrates used for lipid deposition. This can be stated by the fact that mica is a better support for the protein/lipid system.
Also the thermodynamic state of the lipid layer plays a role for protein recrystallization. The clearest case was the existence of regular S-layer structures on condensed DPPC monolayers at 25ºC, while no S-protein adsorption took place on DPPC vesicles.

Although we have found experimental conditions leading to the building of a macromolecular system composed of phospholipids and S-proteins, still very little is known about the recrystallization of different S-proteins on supported flat lipid layers and vesicles. More basic research has to be carried out to understand and control the formation of stable S-layers on lipid supports by improving experimental conditions (lipid ratio mixture, temperature, solvent conditions etc.). This is an important issue to develop in the near future for potential applications of S-layer (fusion) proteins, which will provide different functionalities to the natural lipidic systems.

**Acknowledgements**
This work has been supported by the Spanish Government (grant CTQ2007-66541). MD and SMF thank the financial support of CIC BiomaGUNE (Etortek grant). JLTH thanks the I3 programme of the Spanish Government.